\documentclass[dvips,12pt,a4paper]{article}
\usepackage{a4wide}
\usepackage{epsfig}
\usepackage{amsmath}
\usepackage{latexsym}

  \newlength{\absize}
\newcommand{\dd}{\mbox{{\rm d}}}

\newcommand{\Lumint}{{\cal L}_{\rm int}}

\allowdisplaybreaks 

\catcode`@=11
\def\citer{\@ifnextchar [{\@tempswatrue\@citexr}{\@tempswafalse\@citexr[]}}
 
\def\@citexr[#1]#2{\if@filesw\immediate\write\@auxout{\string\citation{#2}}\fi
  \def\@citea{}\@cite{\@for\@citeb:=#2\do
    {\@citea\def\@citea{--\penalty\@m}\@ifundefined
       {b@\@citeb}{{\bf ?}\@warning
       {Citation `\@citeb' on page \thepage \space undefined}}%
\hbox{\csname b@\@citeb\endcsname}}}{#1}}
\catcode`@=12

\begin{document}
  \thispagestyle{empty}
  \pagestyle{empty}
  \renewcommand{\thefootnote}{\fnsymbol{footnote}}
\newpage\normalsize
    \pagestyle{plain}
    \setlength{\baselineskip}{4ex}\par
    \setcounter{footnote}{0}
    \renewcommand{\thefootnote}{\arabic{footnote}}
\newcommand{\preprint}[1]{%
  \begin{flushright}
    \setlength{\baselineskip}{3ex} #1
  \end{flushright}}
\renewcommand{\title}[1]{%
  \begin{center}
    \LARGE #1
  \end{center}\par}
\renewcommand{\author}[1]{%
  \vspace{2ex}
  {\Large
   \begin{center}
     \setlength{\baselineskip}{3ex} #1 \par
   \end{center}}}
\renewcommand{\thanks}[1]{\footnote{#1}}
\renewcommand{\abstract}[1]{%
  \vspace{2ex}
  \normalsize
  \begin{center}
    \centerline{\bf Abstract}\par
    \vspace{2ex}
    \parbox{\absize}{#1\setlength{\baselineskip}{2.5ex}\par}
  \end{center}}

\begin{flushright}
{\setlength{\baselineskip}{2ex}\par
} 
\end{flushright}
\vspace*{4mm}
\vfill
\title{Model-independent analysis of four-fermion contact interaction at 
$e^+e^-$ collider}
\vfill
\author{
A.A. Pankov 
}
\begin{center}
Technical University of Gomel, 246746 Belarus
\end{center}
\vfill
\abstract
{We study electron-electron contact-interaction searches in the processe  
$e^+e^-\to e^+e^-$ at a future 
$e^+e^-$ Linear Collider with  
both beams longitudinally polarized. 
We evaluate the model-independent constraints 
on the coupling constants, emphasizing the role of beam polarization, 
and make a comparison with the case of $e^+e^-\to\mu^+\mu^-$. 
}
\vspace*{20mm}
\setcounter{footnote}{0}
\vfill

\newpage
    \setcounter{footnote}{0}
    \renewcommand{\thefootnote}{\arabic{footnote}}
    \setcounter{page}{1}

\section{Introduction}
Deviations from the Standard Model (SM) caused by new physics characterized
by very high mass scales $\Lambda$ can systematically be studied at lower
energies by using the effective Lagrangian approach. In this framework,
by integration of the heavy degrees of freedom of the new
theory, an effective Lagrangian which obeys the low energy SM symmetries is
constructed in terms of the SM fields. The resulting interaction consists of
the SM itself as the leading term, plus a series of higher order terms
represented by higher-dimensional local operators that are suppressed by
powers of the scale $\Lambda$. Consequently, the effects
of the new physics can be observed at energies well-below $\Lambda$ as a
deviations from the SM predictions, and can be related to some effective
contact interaction. 
\par 
We consider the effects of the flavor-diagonal, 
helicity conserving, ${eeff}$ contact-interaction 
effective Lagrangian \cite{Eichten1}
\begin{equation}
{\cal L}_{\rm CI}
=\frac{1}{1+\delta_{ef}}\sum_{i,j}g^2_{\rm eff}\hskip
2pt\epsilon_{ij}
\left(\bar e_{i}\gamma_\mu e_{i}\right)
\left(\bar f_{j}\gamma^\mu f_{j}\right),
\label{lagra}
\end{equation}
in the Bhabha scattering process 
\begin{equation}
e^++e^-\to e^++e^-, \label{proc}
\end{equation}
at an $e^+e^-$ Linear Collider (LC) with c.m.\ energy 
$\sqrt s=0.5\hskip 2pt{\rm TeV}$ and polarized electron and positron 
beams \cite{tdr}. In Eq.~(\ref{lagra}): $i,j={\rm L,R}$ denote 
left- or right-handed fermion helicities, $f$ indicates the fermion 
species, so that $\delta_{ef}=1$ for the process (\ref{proc}) under
consideration, and the CI coupling constants are parameterized in terms of 
corresponding mass scales as 
$\epsilon_{ij}={\eta_{ij}}/{{\Lambda^2_{ij}}}$.
Actually, one assumes $g^2_{\rm eff}=4\pi$ to account for the fact that 
the interaction would become strong at $\sqrt s\simeq\Lambda$, 
and by convention 
$\vert\eta_{ij}\vert=\pm 1$ or $\eta_{ij}=0$, leaving the
energy scales $\Lambda_{ij}$ as free, {\it a priori} 
independent parameters.     
\par  
Clearly, at $s\ll\Lambda_{ij}^2$, the Lagrangian (\ref{lagra}) 
can only contribute virtual effects, to be sought for as very small 
deviations of the measured observables from the Standard Model (SM) 
predictions. 
The relative size of such effects are expected to be of order 
$s/\alpha\Lambda^2$, with $\alpha$ the SM coupling (essentially, the fine
structure constant) and, therefore, very high collider energies and 
luminosities are required for this kind of searches. In practice, the 
constraints and the attainable reach on the CI couplings can be numerically 
assessed by comparing the theoretical deviations with the foreseen 
experimental uncertainties on the cross sections. 
\par 
For the case of the Bhabha process (\ref{proc}), the effective 
Lagrangian interaction in Eq.~(\ref{lagra}) envisages the existence of 
six individual, and independent, CI models, contributing to individual 
helicity amplitudes or combinations of them,  
with {\it a priori} free, and nonvanishing, coefficients 
(basically, $\epsilon_{\rm LL},\epsilon_{\rm RR}\ {\rm and}
\ \epsilon_{\rm LR}$ combined with the $\pm$ signs). Correspondingly, 
in principle the most general, and model-independent, analysis of the data 
must account for the situation where all four-fermion effective 
couplings defined in Eq.~(\ref{lagra}) are simultaneously allowed in 
the expression for the cross section. 
Potentially, the different CI couplings may interfere and 
substantially weaken the bounds. Indeed, although the different helicity 
amplitudes by themselves do not interfere, {\it the deviations from the 
SM} could be positive for one helicity amplitude and negative for another, 
so that accidental cancellation might occur in the sought for 
deviations from the SM predictions for the relevant observables. 
\par 
The simplest attitude is to assume non-zero values for only one of the 
couplings (or one specific combination of them) at a time, with all others 
zero, this leads to tests of the specific models mentioned above. 
Also, in many cases, global analyses combining data from different 
experiments relevant to the considered type of coupling are performed. 
Current lower bounds on the corresponding $\Lambda$'s 
obtained along this line from recent analyses of $e^+e^-\to {\bar f}f$ at 
LEP, that include Bhabha scattering, are in the range 8-20 TeV and 
are found to substantially depend on the considered one-parameter scenario 
\cite{Abbaneo:2000nr, Bourilkov}. Examples of results for the $eeff$ 
couplings for the different fermion species in (\ref{lagra}), 
from analyses of different kinds of processes and experiments, can be found, 
{\it e.g.}, in Refs.~\citer{Barger:1998nf,Barger:2000gv}. 
\par 
It should be highly desirable to apply a more general 
(and model-independent) approach to the analysis of 
experimental data, that allows to simultaneously include all 
terms of Eq.~(\ref{lagra}) as independent, non vanishing free 
parameters, and yet to derive separate constraints (or exclusion 
regions) on the values of the CI coupling constants, free from 
potential weakening due to accidental cancellations. 
\par 
Such an analysis is feasible with initial beam longitudinal 
polarization, a possibility envisaged at the LC, 
that allows to extract the individual helicity cross sections from 
suitable combinations of measurable 
polarized cross sections and, consequently, to disentangle the 
constraints on the corresponding CI constants $\epsilon_{ij}$, 
see,{\it e.g.}, Refs.~\citer{Schrempp:1988zy,Babich:2001lm}. 
In what follows, we wish to complement the model-independent analysis of 
$e^+e^-\to{\bar f}f$ with $f\ne e,t$ given in 
Refs.~\cite{Babich:2000kp,Babich:2001lm}, with a discussion of the  
role of the polarized differential cross sections measurable at the 
LC in the derivation of model-independent bounds on the three independent 
four-electron contact interactions relevant to 
the Bhabha process (\ref{proc}).

\section{Observables}
With $P^-$ and $P^+$ the longitudinal polarization  
of the electron and positron beams, respectively, and $\theta$ the angle 
between the incoming and the outgoing electrons in the c.m.\ frame, 
the differential cross section of process (\ref{proc}) at lowest order,
including $\gamma$ and $Z$ exchanges both in the $s$ and $t$ 
channels and the contact interaction (\ref{lagra}), can be written 
in the following form \cite{Schrempp:1988zy,Bardin,Renard}: 
\begin{equation}
\frac{\dd\sigma(P^-,P^+)}{\dd\cos\theta}
=(1-P^-P^+)\,\frac{\dd\sigma_1}{\dd\cos\theta}+
(1+P^-P^+)\,\frac{\dd\sigma_2}{\dd\cos\theta}+
(P^+-P^-)\,\frac{\dd\sigma_P}{\dd\cos\theta}.
\label{cross}
\end{equation}
In Eq.~(\ref{cross}): 
\begin{eqnarray}
\frac{\dd\sigma_1}{\dd\cos\theta}&=&
\frac{\pi\alpha^2}{4s}\left[A_+(1+\cos\theta)^2+A_-(1-\cos\theta)^2\right],
\nonumber \\
\frac{\dd\sigma_2}{\dd\cos\theta}&=&
\frac{\pi\alpha^2}{4s}\, 4A_0, 
\nonumber \\
\frac{\dd\sigma_P}{\dd\cos\theta}&=&
\frac{\pi\alpha^2}{4s}\, A_+^P(1+\cos\theta)^2,
\label{sigP}
\end{eqnarray}
with 
\begin{eqnarray}
A_0(s,t)&=&
\left(\frac{s}{t}\right)^2\big\vert 1+g_{\rm R}\hskip 2pt
g_{\rm L}\chi_Z(t)+\frac{t}{\alpha}\,\epsilon_{\rm LR}\big\vert^2,
\nonumber \\
A_+(s,t)&=&
\frac{1}{2}\big\vert 1+\frac{s}{t}+g_{\rm L}^2\hskip 2pt
\left(\chi_Z(s)+\frac{s}{t}\,\chi_Z(t)\right)+
2\frac{s}{\alpha}\,\epsilon_{\rm LL}\big\vert^2 \nonumber \\
&+&
\frac{1}{2}\big\vert 1+\frac{s}{t}+g_{\rm R}^2\hskip 2pt
\left(\chi_Z(s)+\frac{s}{t}\,\chi_Z(t)\right)+
2\frac{s}{\alpha}\,\epsilon_{\rm RR}\big\vert^2, \nonumber \\
A_-(s,t)&=&
\big\vert 1+g_{\rm R}\hskip 2pt
g_{\rm L}\,\chi_Z(s)+\frac{s}{\alpha}\,\epsilon_{\rm LR}\big\vert^2, 
\nonumber \\
A_+^P(s,t)&=&
\frac{1}{2}\big\vert 1+\frac{s}{t}+g_{\rm L}^2\hskip 2pt
\left(\chi_Z(s)+\frac{s}{t}\,\chi_Z(t)\right)+
2\frac{s}{\alpha}\,\epsilon_{\rm LL}\big\vert^2 \nonumber \\
&-&
\frac{1}{2}\big\vert 1+\frac{s}{t}+g_{\rm R}^2\hskip 2pt
\left(\chi_Z(s)+\frac{s}{t}\,\chi_Z(t)\right)+
2\frac{s}{\alpha}\,\epsilon_{\rm RR}\big\vert^2.
\label{A}
\end{eqnarray}
Here: $\alpha$ is the fine structure constant; $t=-s(1-\cos\theta)/2$ and 
$\chi_Z(s)=s/(s-M^2_Z+iM_Z\Gamma_Z)$,  
$\chi_Z(t)=t/(t-M^2_Z)$ represent the $Z$ propagator in the $s$ and $t$ 
channels,
respectively, with $M_Z$ and $\Gamma_Z$ the mass and width of the $Z$;  
$g_{\rm R}=\tan\theta_W$, $g_{\rm L}=-\cot{2\,\theta_W}$
are the SM right- and left-handed electron couplings of the $Z$, 
with $\theta_W$ the electroweak mixing angle.
\par
With both beams polarized, the polarization of each beam can be changed 
on a pulse by pulse basis. This would allow the separate measurement 
of the polarized cross sections for each of the four polarization 
configurations $++$, $--$, $+-$ and $-+$, corresponding to the four sets 
of beam polarizations 
$(P^-,P^+)=(P_1,P_2)$, $(-P_1,-P_2)$, $(P_1,-P_2)$ and $(-P_1,P_2)$, 
respectively, with $P_{1,2}>0$. Specifically, with the simplifying 
notation $\dd\sigma\equiv\dd\sigma/\dd\cos\theta$:
\begin{eqnarray}
{\dd\sigma_{++}}&\equiv&{\dd\sigma(P_1,P_2)}=
(1-P_1P_2)\,{\dd\sigma_1}+
(1+P_1P_2)\,{\dd\sigma_2}+(P_2-P_1)\,{\dd\sigma_P},
\nonumber \\
{\dd\sigma_{--}}&\equiv&{\dd\sigma(-P_1,-P_2)}=
(1-P_1P_2)\,{\dd\sigma_1}+
(1+P_1P_2)\,{\dd\sigma_2}-(P_2-P_1)\,{\dd\sigma_P},
\nonumber \\
{\dd\sigma_{+-}}&\equiv&{\dd\sigma(P_1,-P_2)}=
(1+P_1P_2)\,{\dd\sigma_1}+
(1-P_1P_2)\,{\dd\sigma_2}-(P_2+P_1)\,{\dd\sigma_P},
\nonumber \\
{\dd\sigma_{-+}}&\equiv&{\dd\sigma(-P_1,P_2)}=
(1+P_1P_2)\,{\dd\sigma_1}+
(1-P_1P_2)\,{\dd\sigma_2}+(P_2+P_1)\,{\dd\sigma_P}.
\label{cross4}
\end{eqnarray}
To extract from the measured polarized cross sections the values of 
$\dd\sigma_1$, $\dd\sigma_2$ and $\dd\sigma_P$,  
that carry the information on the CI couplings, one has to invert the 
system of equations (\ref{cross4}). The 
solution reads:
\begin{eqnarray}
{\dd\sigma_1}&=&\frac{1}{8}\,\left[(1-\frac{1}{P_1P_2})\,
({\dd\sigma_{++}}+{\dd\sigma_{--}})+
(1+\frac{1}{P_1P_2})\,
({\dd\sigma_{+-}}+{\dd\sigma_{-+}})\right],
\nonumber \\
{\dd\sigma_2}&=&\frac{1}{8}\,\left[(1+\frac{1}{P_1P_2})\,
({\dd\sigma_{++}}+{\dd\sigma_{--}})+
(1-\frac{1}{P_1P_2})\,
({\dd\sigma_{+-}}+{\dd\sigma_{-+}})\right],
\nonumber \\
{\dd\sigma_P}&=&-\frac{1}{2\,(P_1+P_2)}\,
\left({\dd\sigma_{+-}}-{\dd\sigma_{-+}}\right)=
\frac{1}{2\,(P_2-P_1)}\,
\left({\dd\sigma_{++}}-{\dd\sigma_{--}}\right).
\label{observ}
\end{eqnarray}
Notice that the equations in (\ref{cross4}) are not all 
linearly independent, and that not only $P_{1}\ne 0$ and $P_{2}\ne 0$, but  
also $P_1\ne P_2$ is needed to obtain Eqs.~(\ref{observ}). As one 
can see from Eqs.(\ref{sigP}) and (\ref{A}), $\sigma_2$ depends on 
only one contact interaction parameter ($\epsilon_{\rm LR}$), $\sigma_P$ 
is two-parameter dependent ($\epsilon_{\rm RR}$ and $\epsilon_{\rm LL}$), 
and $\sigma_1$ depends on all three parameters. Therefore, the derivation 
of the model-independent constraints on the CI couplings requires the 
combination all polarized observables of Eq.~(\ref{observ}). In this 
regard, to emphasize the role of polarization, one can observe from 
Eqs.~(\ref{cross})-(\ref{A}) that in the unpolarized case $P_1=P_2=0$, 
where only 
$\sigma_1$ and $\sigma_2$ appear, the interference of the 
$\epsilon_{\rm LR}$ term with the SM amplitude in 
$A_0$ and $A_-$ has opposite signs, leading to a partial cancellation 
for $-t\sim s$. Consequently, as briefly anticipated in Sect.~1, one 
expects the unpolarized cross section to have reduced sensitivity to 
$\epsilon_{\rm LR}$. Conversely, $\epsilon_{\rm LR}$ is {\it directly} 
accessible from $\dd\sigma_2$, {\it via} polarized cross sections as 
in Eq.~(\ref{observ}). Also, considering that numerically 
$g_{\rm L}^2\cong g_{\rm R}^2$, the parameters $\epsilon_{\rm LL}$ and 
$\epsilon_{\rm RR}$ contribute to the unpolarized cross section through 
$A_+$ with equal coefficients, so that, in general, only correlations of the 
form $\vert\epsilon_{\rm LL}+\epsilon_{\rm RR}\vert<{\rm const}$, 
and not finite allowed regions, could be derived in this case. 
\par  
To make contact to the experiment we take $P_1=0.8$ and $P_2=0.6$, and 
impose a cut in the forward and backward directions. Specifically, 
we consider the cut angular range $\vert\cos\theta\vert<0.9$ and divide 
it into nine equal-size bins of width $\Delta z=0.2$ ($z\equiv\cos\theta$). 
We also introduce the experimental efficiency, $\epsilon$, for
detecting the final $e^+e^-$ pair and, according to the LEP2 experience,  
$\epsilon=0.9$ is assumed. 
\par 
We then define the four, directly measurable, event rates integrated 
over each bin:
\begin{equation}
N_{++},\quad  N_{--},\quad N_{+-},\quad N_{-+},
\label{obsn}
\end{equation}
and ($\alpha\beta=++$, etc.):
\begin{equation}
N_{\alpha\beta}^{\rm bin}=\frac{1}{4}\Lumint\,\epsilon
\int_{\rm bin}(\dd\sigma_{\alpha\beta}/\dd z)\dd z.
\label{n}
\end{equation}
In Eq.~(\ref{n}), $\Lumint$ is the time-integrated luminosity, which is 
assumed to be equally divided among over the four combinations of 
electron and positron beams polarization defined in Eqs.~(\ref{cross4}). 
\par 
In Fig.~1, the bin-integrated angular distributions of
$N_{++}^{\rm bin}$ and $N_{+-}^{\rm bin}$ in the SM 
at $\sqrt{s}=500$ GeV and  
$\Lumint=50\ \mbox{fb}^{-1}$ are presented as histograms. Here, the SM 
cross sections have been evaluated by means of the effective 
Born approximation \cite{Consoli:1989pc,Altarelli:1990dt}.
The typical forward peak, dominated by the $t$-channel photon pole,  
dramatically shows up, and determines a really large statistics 
available in the region of small $t$. The $\cos\theta$ distributions for 
the other polarization configurations in (\ref{cross4}) are similar 
and, therefore, we do not represent them here.  

\begin{figure}[thb]
\refstepcounter{figure}
\label{Fig1}
\addtocounter{figure}{-1}
\begin{center}
\setlength{\unitlength}{1cm}
\begin{picture}(8.5,7.8)
\put(-1.0,-1.5){
\mbox{\epsfysize=10.0cm\epsffile{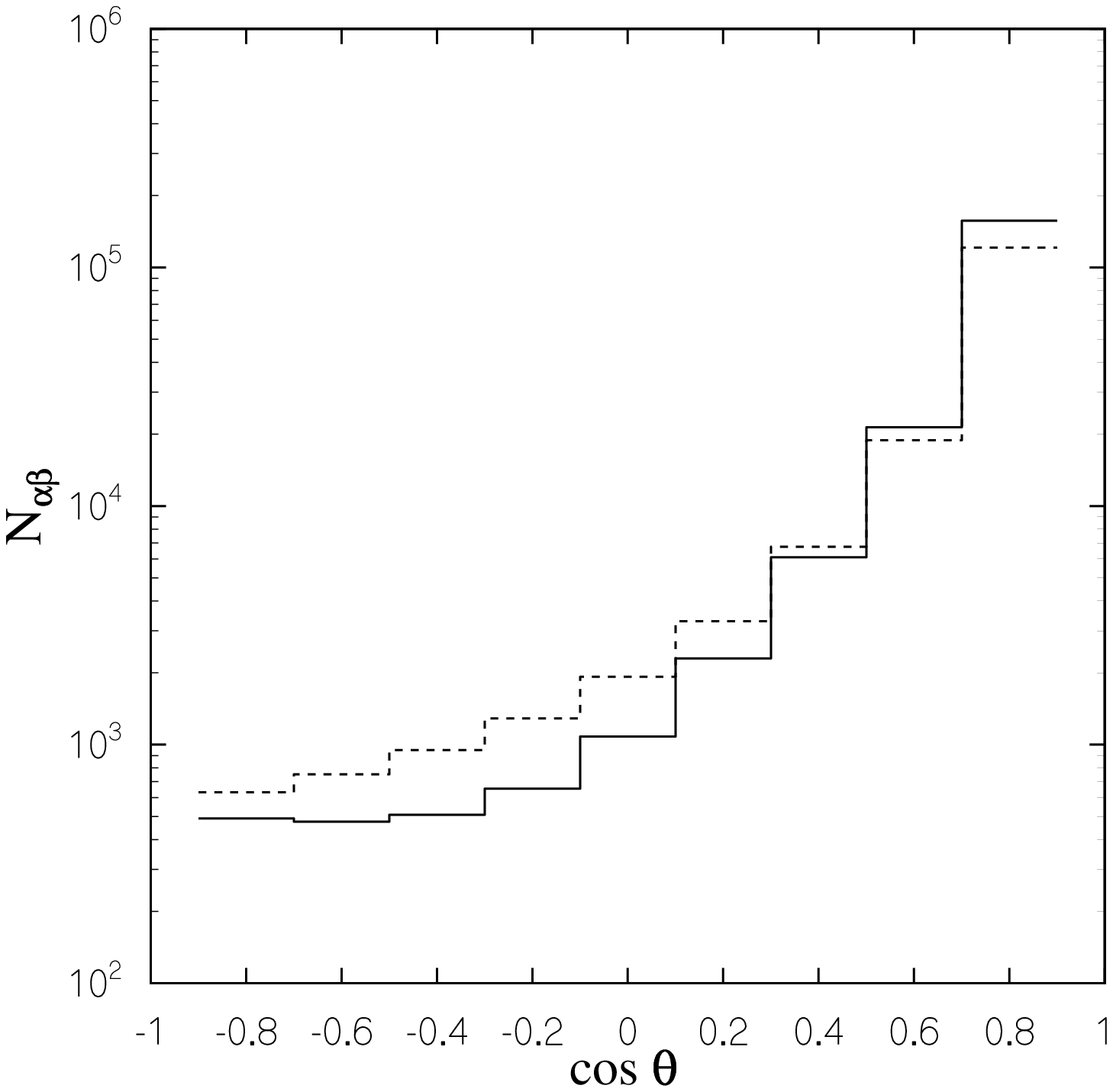}}}
\end{picture}
\vspace*{2mm}
\caption{Bin-integrated angular distributions of
$N_{+-}^{\rm bin}$ (solid line) and $N_{++}^{\rm bin}$ (dashed line), 
Eq.(\ref{n}), in the SM at $\sqrt{s}=500$ GeV  and 
$\Lumint=50\ \mbox{fb}^{-1}$.
}
\end{center}
\end{figure}
\par 
The next step is to define the relative deviations of the 
cross sections $\sigma_1$, $\sigma_2$ and $\sigma_P$ from the SM 
predictions, due to the contact interaction. In general, for such 
deviations, we use the notation: 
\begin{equation}
\Delta_{\cal O}=\frac{{\cal O}(SM+CI)-{\cal O}(SM)}{{\cal O}(SM)},
\label{relat}
\end{equation}  
To get an illustration of the effect of the contact interactions on the 
observables (\ref{observ}) under consideration, we show in 
Fig.~2a and Fig.~2b the angular distributions of the relative deviations of 
$\dd\sigma_1$ and $\dd\sigma_2$, taking as examples the values of 
$\Lumint$ and $\Lambda_{ij}$ indicated in the captions. 
The SM predictions are 
evaluated in the same, effective Born, approximation as in Fig.~1. 
The deviations are 
then compared to the expected statistical uncertainties, represented by 
the vertical bars. Fig.~2a shows that $\dd\sigma_1$ is sensitive to 
contact interactions in the forward region, where the ratio of the 
`signal' to the statistical uncertainty increases. Also, it indicates 
that, for the chosen values of the c.m. energy $\sqrt s$ and $\Lumint$, 
the reach on $\Lambda_{ij}$ will be substantially larger than
30 TeV. Conversely, Fig.~2b shows that the sensitivity of $\dd\sigma_2$ 
is almost independent on the chosen kinematical range in $\cos\theta$, 
leading to a really high sensitivity of this observable to 
$\epsilon_{\rm LR}$, and to corresponding lower bounds on $\Lambda_{\rm LR}$ 
potentially larger than 50 TeV. 
\begin{figure}[htb]
\refstepcounter{figure}
\label{Fig2}
\addtocounter{figure}{-1}
\begin{center}
\setlength{\unitlength}{1cm}
\begin{picture}(12,7.5)
\put(-3.,0.0)
{\mbox{\epsfysize=8.5cm\epsffile{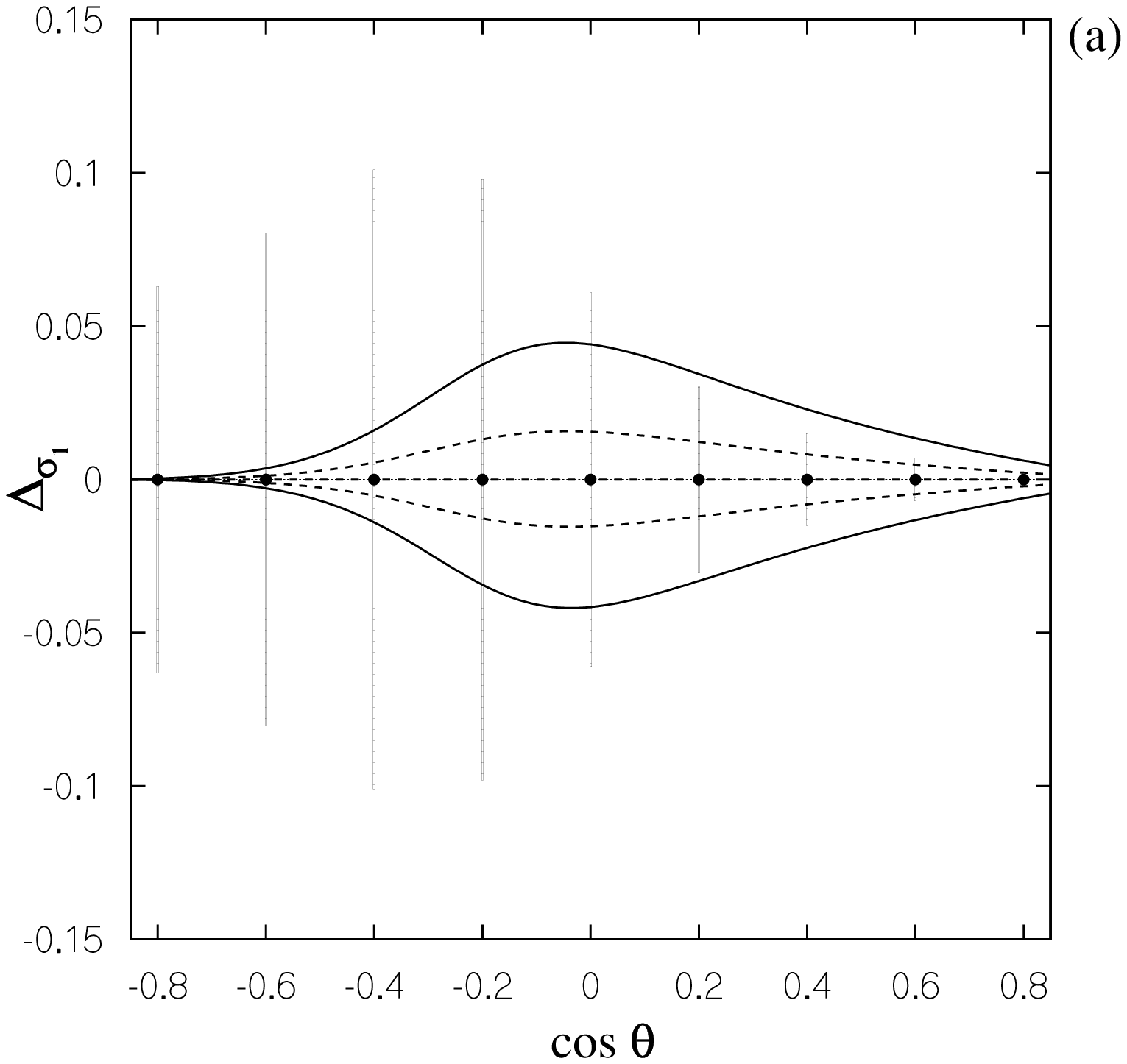}}
 \mbox{\epsfysize=8.5cm\epsffile{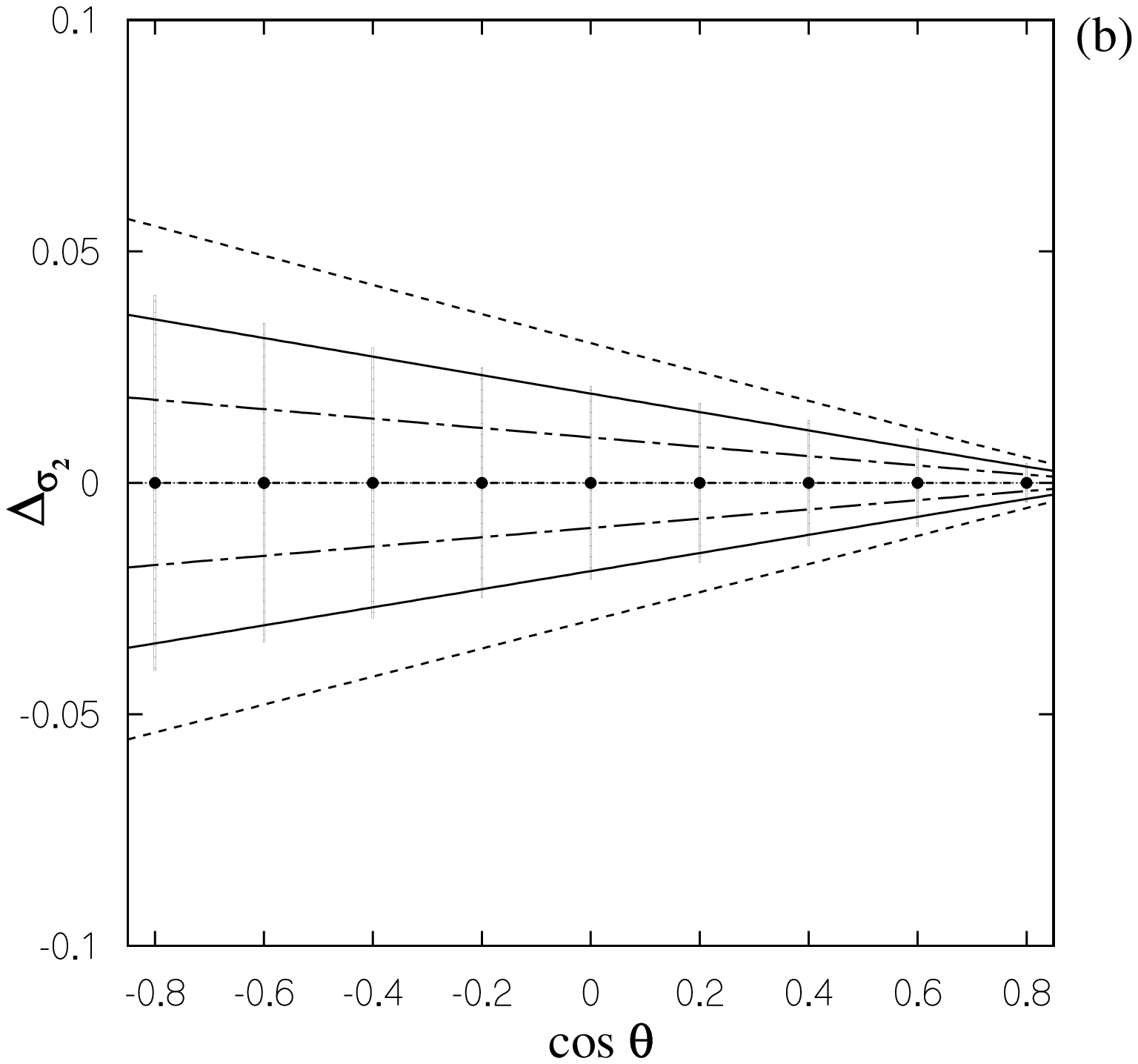}}}
\end{picture}
\vspace*{-13mm}
\caption{The angular distributions for the relative deviations 
$\Delta_{\sigma_1}$  (a) at $\Lambda_{\rm RR}$=30 TeV (solid line) and 
50 TeV (dashed line), and for $\Delta_{\sigma_2}$ (b) at 
$\Lambda_{\rm LR}$=40 TeV (dashed line), 50 TeV (solid line),
70 TeV (dot-dashed line). The curves above (below) the horizontal line 
correspond to negative (positive) interference between contact interaction 
and SM amplitude.  The error bars show the expected statistical error 
at $\Lumint=50\ \mbox{fb}^{-1}$.}
\end{center}
\end{figure}
We now proceed to the analysis of the bounds on the contact interaction 
couplings.

\section{Constraints on CI couplings}

To assess the sensitivity to the compositeness scale we assume the data 
to be well described by the SM predictions ($\epsilon_{ij}=0$), 
{\it i.e.,} that no deviation is observed within the foreseen experimental 
accuracy, and perform a $\chi^2$ analysis of the $\cos\theta$ angular 
distribution. For each of the observable cross sections, 
the $\chi^2$ distribution is defined as the sum over the above mentioned 
nine equal-size $\cos\theta$ bins:
\begin{equation}
\chi^2({\cal O})=
\sum_{\rm bins}\left(\frac{\Delta{\cal O}^{\rm bin}}
{\delta{\cal O}^{\rm bin}}\right)^2, 
\label{chi}
\end{equation}
where ${\cal O}=\sigma_1$, $\sigma_2$, $\sigma_P$ and 
$\sigma^{\rm bin}\equiv\int_{\rm bin}(\dd\sigma/\dd z)\dd z$. In 
Eq.~(\ref{chi}), $\Delta{\cal O}$ represents the deviation from the 
SM prediction, $\Delta{\cal O}={\cal O}(SM+CI)-{\cal O}(SM)$, 
that can be easily expressed in terms of the CI couplings by using 
Eqs.~(\ref{A}), and $\delta{\cal O}$ is the expected experimental 
uncertainty, that combines the statistical and the systematic one.
\par 
In the following analysis, the theoretical expectations for the polarized 
cross sections are evaluated by using the program TOPAZ0 
\cite{Montagna1, Montagna2}, adapted to the present discussion, 
with $m_{\rm top}=175$ GeV and $M_H=120$ GeV. For electron-positron 
final states, a cut on the acollinearity angle between electron and 
positron, $\theta_{\rm acol}<10^\circ$, is applied to select non-radiative 
events.
\par
Concerning the numerical inputs and assumptions used in the estimate of 
$\delta\cal O$, to assess the role of statistics we vary $\Lumint$ 
from $50$ to $500\ \mbox{fb}^{-1}$ (a quarter of total the running time 
for each polarization configuration). As for the systematic 
uncertainty, we take $\delta\Lumint/\Lumint=0.5\%$, 
$\delta\epsilon/\epsilon=0.5\%$ and, regarding the electron and positron 
degrees of polarization, $\delta P_1/P_1=\delta P_2/P_2=0.5\ \%$.
\par 
As a criterion to constrain the allowed values of the contact interaction
parameters by the non-observation of the corresponding deviations, 
we impose $\chi^2<\chi^2_{\rm CL}$, where the actual
value of $\chi^2_{\rm CL}$ specifies the desired `confidence' level.
We take the values $\chi^2_{\rm CL}=$3.84 and 5.99 for 95\% C.L. for 
a one- and a two-parameter fit, respectively.
\par 
We begin the presentation of the numerical results from the 
consideration of $\epsilon_{\rm LR}$. For this case, the relevant 
cross section $\sigma_2$ depends on $\epsilon_{\rm LR}$ only, 
see Eqs.~(\ref{sigP}) and (\ref{A}) and, therefore, the constraints on 
that parameter are determined from a one-parameter fit. 
The model-independent, discovery reach expected at the LC for the 
corresponding mass scale $\Lambda_{\rm LR}$ is represented, as a 
function of the integrated luminosity ${\Lumint}$, by the solid line 
in Fig.~3. As expected, the highest luminosity determines the strongest  
constraints on the CI couplings.\footnote{Such increase with luminosity 
is somewhat slower than expected from the scaling law 
$\Lambda\sim\left(s\Lumint\right)^{1/4}$, as the effect of the systematic 
uncertainties competing with the statistical ones.} Fig.~3 dramatically 
shows the really high sensitivity of $\sigma_2$, such that the discovery 
limits on $\Lambda_{\rm LR}$ are the highest, compared to the 
$\Lambda_{\rm RR}$ and $\Lambda_{\rm LL}$ cases, and can be as large 
as 110 up to 170 times the total c.m. energy. 

\begin{figure}[thb]
\refstepcounter{figure}
\label{Fig3}
\addtocounter{figure}{-1}
\begin{center}
\setlength{\unitlength}{1cm}
\begin{picture}(8.5,7.8)
\put(-1.0,-1.5){
\mbox{\epsfysize=10.0cm\epsffile{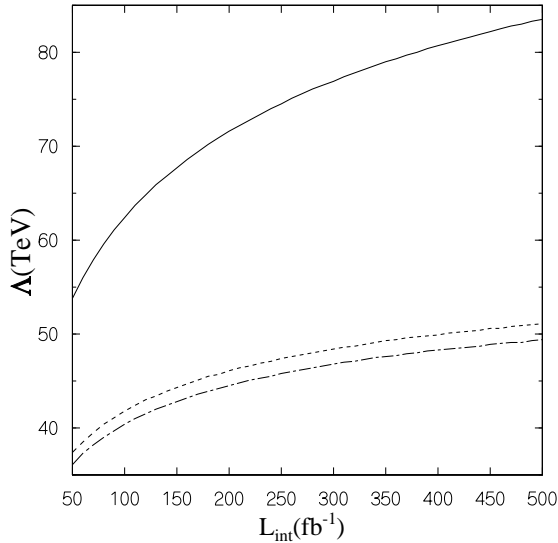}}}
\end{picture}
\vspace*{2mm}
\caption{
Reach in $\Lambda_{ij}$ at 95\% C.L. {\it vs.} integrated 
luminosity ${\Lumint}$ obtained from the model-independent analysis
for $e^++e^-\to e^++e^-$ at $E_{\rm c.m.}=0.5$~TeV,
$\vert P^-\vert=0.8$ and $\vert P^+\vert=0.6$, $\Lambda_{\rm LR}$ (solid line),
$\Lambda_{\rm LL}$ (dashed line), $\Lambda_{\rm RR}$ (dot-dash line).
}
\end{center}
\end{figure}
\par
Since $\sigma_P$ simultaneously depends on the two independent 
CI couplings $\epsilon_{\rm RR}$ and $\epsilon_{\rm LL}$, a 
two-parameter analysis is needed in this case. Since the terms 
quadratic in $\epsilon_{\rm LL}$ and $\epsilon_{\rm RR}$ largely cancel 
leaving the remaining interference to dominate the 
relevant deviations from the SM, the resulting constraint has the 
form of a straight band, as depicted in Fig.~4a. Indeed, such a 
band represents a correlation between the two parameters, rather than 
a bound around the SM value $\epsilon_{\rm LL}=\epsilon_{\rm RR}=0$. 
\par
In order to get a restricted allowed region around zero, one can combine 
the band with the exclusion region obtained from $\sigma_1$. 
However, since the latter depends on {\it all three} contact interaction 
parameters, see Eqs.~(\ref{sigP}) and (\ref{A}), to set constraints in the 
($\epsilon_{\rm RR},\epsilon_{\rm LL}$) plane requires the combination 
of the $\sigma_1$-bounds with the limits on $\epsilon_{\rm LR}$ 
derived above from $\sigma_2$. The bound in the 
($\epsilon_{\rm RR},\epsilon_{\rm LL}$) resulting from this procedure 
is shown in Fig.~4a and, finally, the shaded ellipse determined by the 
combination with the band determined by $\sigma_P$ represents the 
restricted allowed region around the SM point $\epsilon_{ij}=0$. 
With reference to Eq.~(\ref{chi}), for the $\chi^2$ 
analysis this amounts to the consideration of the combined 
$\chi^2(\sigma_1)+\chi^2(\sigma_P)$. 
Fig.~4b is essentially a magnification of the shaded region of 
Fig.~4a, and represents the model-independent limits  
on $\epsilon_{\rm LL}$ and $\epsilon_{\rm RR}$ attainable at the 
considered LC, for two possible values of the integrated luminosity. 
These bounds are translated into the model-independent reach on the 
mass scale parameters $\Lambda_{\rm LL}$ and $\Lambda_{\rm RR}$, 
represented as a function of luminosity in Fig.~3. The fact that 
such bounds are substantially lower than those for $\Lambda_{\rm LR}$ 
reflects that a combined two-parameter $\chi^2$ analysis must be used. 
In this regard, the calculation presented here indicates that not only 
polarization,  but also combinations of measurements of polarized 
observables are necessary to obtain model-independent bounds on the 
CI couplings. 
 
\begin{figure}[htb]
\refstepcounter{figure}
\label{Fig4}
\addtocounter{figure}{-1}
\begin{center}
\setlength{\unitlength}{1cm}
\begin{picture}(12,7.5)
\put(-3.,0.0)
{\mbox{\epsfysize=8.5cm\epsffile{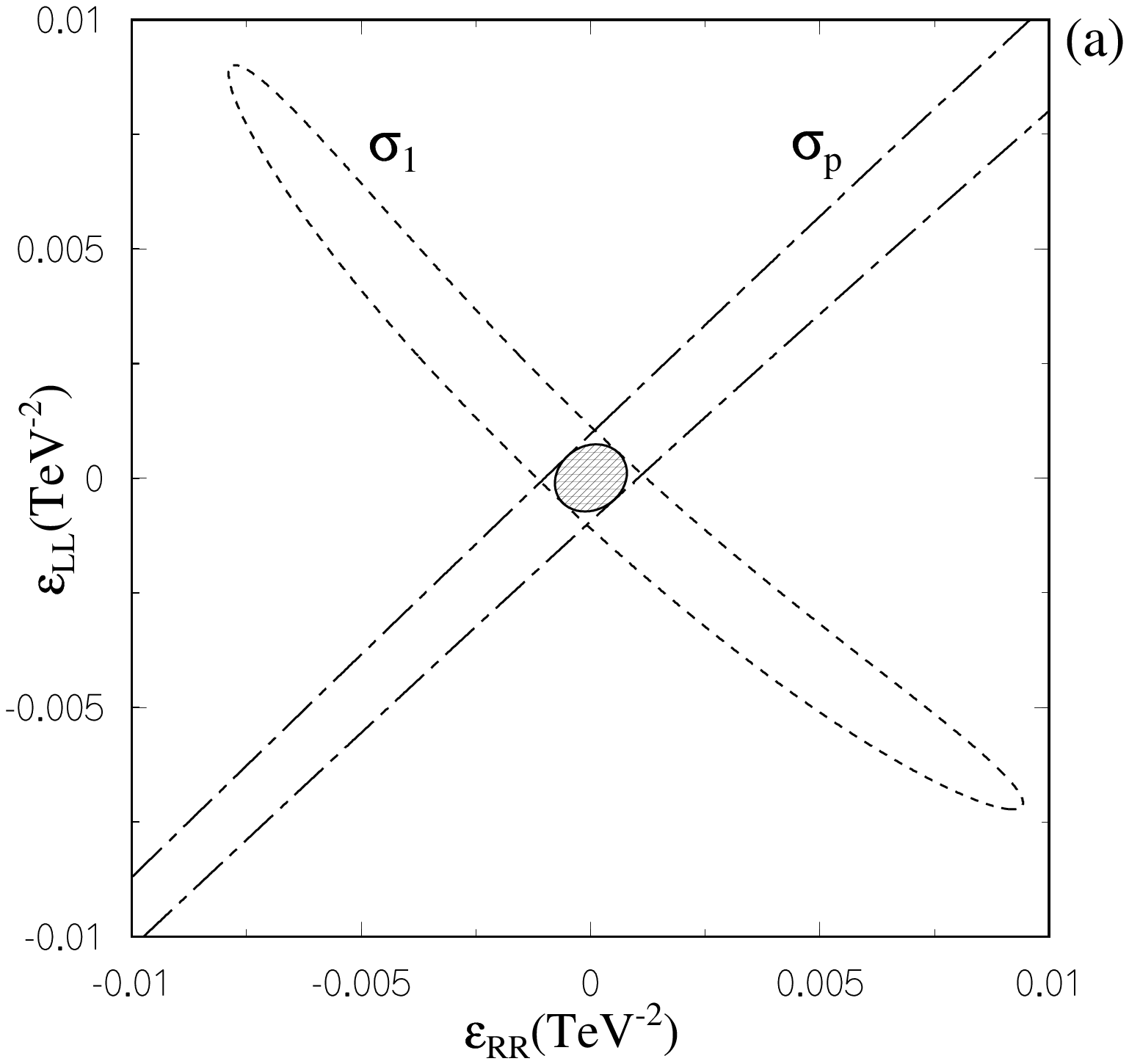}}
 \mbox{\epsfysize=8.5cm\epsffile{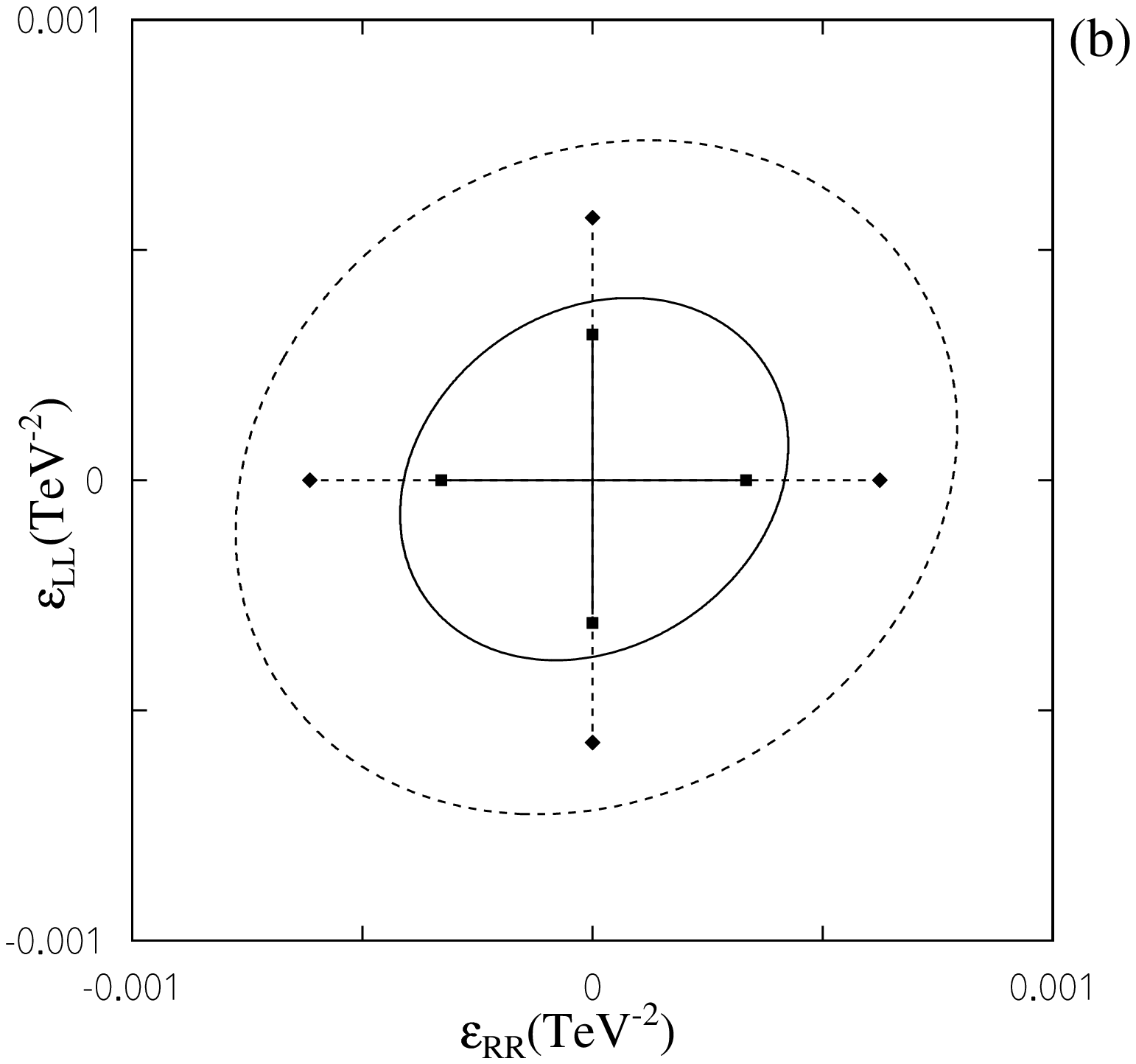}}}
\end{picture}
\vspace*{-13mm}
\caption{
(a) Allowed areas at 95\% C.L. on electron contact interaction parameters
in the planes ($\epsilon_{\rm RR},\epsilon_{\rm LL}$), 
obtained from $\sigma_1$ and $\sigma_P$ at $\sqrt{s}=500$ GeV and 
$\Lumint=50\ \mbox{fb}^{-1}$(a); 
(b) Combined allowed regions at 95\% C.L. obtained from $\sigma_1$ and 
$\sigma_P$ at $\sqrt{s}=500$ GeV, $\Lumint=50\ \mbox{fb}^{-1}$ (outer ellipse)
and $500\ \mbox{fb}^{-1}$ (inner ellipse).}
\end{center}
\end{figure}


The crosses in Fig.~4b represent the model-dependent constraints 
obtainable by taking only one non-zero parameter at a time, instead 
of two simultaneously non-zero, and independent, as in the analysis 
discussed above. Similar to the inner and outer ellipses, the shorter 
and longer arms of the crosses refer to integrated luminosity 
$\Lumint=50\ \mbox{fb}^{-1}$ and $500\ \mbox{fb}^{-1}$, respectively.  
One can note from Fig.~4b that the `single-parameter' constraints on 
the individual CI parameters $\epsilon_{\rm RR}$ and $\epsilon_{\rm LL}$
are numerically more stringent, as compared to the 
model-independent ones. Essentially, this is a reflection of the smaller 
value of the critical $\chi^2$, $\chi^2_{\rm crit}=3.84$ corresponding to 
95\% C.L. with a {\it one-parameter} fit, and also of a reduced role of 
correlations among the different observables. 

\section{Concluding remarks}
In the previous sections we have derived limits on the contact interactions  
relevant to Bhabha scattering by a model-independent analysis 
that allows to simultaneously account for all independent couplings as 
non-vanishing free parameters. The results for the lower bounds on the 
corresponding mass scales $\Lambda$ range, depending on the luminosity, 
from essentially 38 to 50 TeV for the LL and RR cases, and from 54 to 
84 TeV for the LR case. The comparison with the numerical results relevant 
to the $e^+e^-\to\mu^+\mu^-$ channel, derived from a similar analysis 
\cite{Babich:2001lm}, is shown in Table 1.   

\begin{table}[ht]
\centering
\caption{
Reach in $\Lambda_{ij}$ at 95\% C.L., from the
model-independent analysis performed 
for $e^+e^-\to\mu^+\mu^-$ and $e^+e^-$, 
at $E_{\rm c.m.}=0.5$~TeV, $\Lumint=50\,\mbox{fb}^{-1}$ and
$500\,\mbox{fb}^{-1}$, $\vert P^-\vert=0.8$ and 
$\vert P^+\vert=0.6$.
}
\medskip
{\renewcommand{\arraystretch}{1.2}
\begin{tabular}{|c|c|c|c|c|c|}
\hline
${\rm process}$ & ${\Lumint}$ & $\Lambda_{\rm LL}$ & $\Lambda_{\rm RR}$ &
$\Lambda_{\rm LR}$ & $\Lambda_{\rm RL}$ \\
&$\mbox{fb}^{-1}$& TeV & TeV &TeV  & TeV \\
\hline
\cline{2-6}
 & 50 & $ 35 $ & $ 35 $ & $ 31 $ & $ 31 $
\\
\cline{2-6}
$e^+e^-\to\mu^+\mu^-$
 & 500 &$ 47 $ & $ 49 $ & $ 51 $ & $ 52 $ 
\\ \hline
\cline{2-6}
 & 50 & $ 38 $ & $ 36 $ & $ 54 $ &
\\
\cline{2-6}
$e^+e^-\to e^+ e^-$
 & 500 & $ 51 $ & $ 49 $ & $ 84 $ &  
\\ \hline
\end{tabular}
} 
\label{tab:table-1}
\end{table}

The table shows that for $\Lambda_{\rm LL}$ and $\Lambda_{\rm RR}$ the
restrictions from $e^+e^-\to\mu^+\mu^-$ and $e^+e^-\to e^+e^-$ are 
qualitatively comparable. Instead, the sensitivity to $\Lambda_{\rm LR}$, 
and the corresponding lower bound, is dramatically higher in the 
case of Bhabha scattering. In this regard, this is the consequence of the 
initial beams longitudinal polarization that allows, by 
measuring suitable combinations of polarized cross sections, to directly 
disentangle the coupling $\epsilon_{\rm LR}$. Indeed, without polarization, 
as previously observed, in general only correlations among couplings, 
rather that finite allowed regions, could be derived. Alternatively, in 
the unpolarized case, a one-parameter analysis testing individual models 
can be performed.

\medskip
\leftline{\bf Acknowledgements}
\par\noindent

I would like to thank Prof. N. Paver for the enjoyable collaboration on the subject
matter covered here.

\goodbreak


\end{document}